# A Novel Study of the Relation Between Students' Navigational Behavior on Blackboard and their Learning Performance in an Undergraduate Networking Course

**Abstract**

This paper provides an overview of students' behavior analysis on a learning management system (LMS), Blackboard (Bb) Learn for a core data communications course of the Undergraduate IT program in the Information Sciences and Technology (IST) Department at George Mason University (GMU). This study is an attempt to understand the navigational behavior of students on Blackboard Learn which can be further attributed to the overall performance of the students. In total, 160 undergraduate students participated in the study. Vast amount of student's activities data across all four sections of the course were collected. All sections have similar content, assessment design and instruction methods. A correlation analysis between the different assessment methods and various key variables such as total student time, total number of logins and various other factors were performed, to evaluate students' engagement on Blackboard Learn. Our findings can help instructors to efficiently identify students' strengths or weaknesses and fine-tune their courses for better student engagement and performance.

## 1. Introduction

The evolution of learning management system (LMS), Blackboard (Bb) as a virtual learning environment and course management system had a significant impact on education. Bb has powered online education and offered stability and scalability for wider institutional application. Since the initial release of Bb, various analytical tools have been increased for teachers to evaluate students' navigational behaviors. With the advent of strong computational resources, now it is more feasible to compute and analyze those behaviors, which was not possible two decades ago.

It has been noticed that difficult undergraduate mandatory courses struggle with attrition issues and this situation is quite prevalent in other universities as well [3]. Despite numerous efforts from educators, we still observe that many talented and keen students still encounter problems to ace such courses. During the last decade, there has been extensive research to align courses according to industry environments, and we at GMU-IST department have conducted some relevant research studies [1] on how we can line up assessment methods with learning objectives. Black and William [2] have used the term "assessment" as a method employed by teachers to evaluate students that also serve as feedback to the instructors.

Some educators reason that students may have the required skills but they perform poorly due to lack of management skills or commitment required to excel in the networking courses. This paper focuses on navigational behavior data collected from Bb to come up with noteworthy factors that can help us identify the reasons behind student failure or success. Furthermore, we will use these factors to better monitor and supervise the students to help them succeed in their courses. This study will also enable teachers to skillfully manage their course content on Bb.

For this study, we are using the Bb data of 4 networking course sections. At GMU, Bb is the official LMS for course content, assignments, discussion forums, quizzes and labs. Instructors use Bb to upload course content such as course materials, conduct quizzes and communicate with students. Students review the course material and submit their assignments, labs and quizzes to Blackboard. This study will help us to come up with a better understanding and strategy to enhance student learning performance.

## 2. Literature Review

A number of researchers have studied student habits on how they work on different assessment methods. They have tried to analyze how students work on their assignments. One such study is analyzed by Edwards et al. [3] in which he discusses the behavior of students as they attempt programming assignments. The study was conducted using a large data set of 89,879 assignments submitted by 1101 students. To ensure a fair comparison, students who consistently received A/B and C/D/F course grades were removed from the dataset. They highlighted significant factors; when students started assignments early and submitted before deadline, they scored better than those who started late and submitted before deadline. It can be due to the fact that students who start earlier have more time to seek guidance.

Fenwick et al. [4] carried out an important study which examined the behaviors of novice programmers in introductory programming courses. This study encompasses CS 1 and CS 2 students from fall and spring semesters. In [5], the authors collected and analyzed data from each student's machine, and monitored the students' compilation habits. The article [4] concluded that students who started two or more days prior to the deadline earned a better grade than those who started a day before the deadline. Similarly, Retina [6] is another tool that can be used in conjunction with Interactive Development Environment (IDE) for observing the behavior of students as they progress through their assignments. Murphy et al. [6] study included only 21 students and concluded that students who spent less time on the assignments showed to do better than students who spent more time He also concluded students generally underestimate the time taken by the assignments.

Another similar study was carried out by Mierle, et al. [7] to find key predictors for providing assistance to those students who are struggling through the undergraduate computer course. For this study, they analyzed the Concurrent Versions System (CVS) repositories of over 200 students, and concluded results which are dissimilar to other studies. Their results suggested that students' behavior is not very much correlated with their performance, as long as they submit the assignments. Furthermore, they came to the conclusion that there is not a significant difference in students who started well before the deadline than those who started late.

In [8], "Blackbox: A Large Scale Repository of Novice Programmers' Activity" discusses about collection of large scale data. The anonymous data from over one hundred thousand users is available to researchers for use in their respective studies. The author discusses the types of analysis that can be carried out using the data available via Blackbox, and the challenges that lie ahead in automated analytics. Brown et al. [8] also discusses how this data can be used to track the student behavior and error count replication. Similarly, Jadud [9] was the first one to carry out a research study to discuss the association of student behavior with the frequency and types of error incurred.

All the prior studies have focused mostly on programming assignments and tried to figure out how students' behavior impacts their performance. Some authors just focused on the type of errors incurred by the students. However, the results of these studies are not conclusive. Some of them don't have enough profundity for tracking students' behavior throughout the course. In contrast, our study is not limited to just one type of assignment but it covers the student behavior all across the course such as lecture slides, lab assignments, home assignments, skills assessments and quizzes. Furthermore, we are not only trying to find a relation but also, we have developed a model that can predict the student performance based on the selected key features. This study will also help instructors to better manage the course content.

Table 1: Data Set Snippet

| Student ID | Sunday | Monday | ... | Saturday | Total Time in Course | Total Items accessed | Total Logins | Time in Course Content | ... | Score |
|---|---|---|---|---|---|---|---|---|---|---|
| 53433 | 1.95 | 9.13 | | 0.56 | 16.59 | 30 | 100 | 0.06 | | 63.5 |
| 31231 | 2.65 | 8.31 | | 0.16 | 15 | 31 | 142 | 0.03 | | 92 |
| 31231 | 0.94 | 7.24 | | 0 | 11.48 | 29 | 131 | 0.34 | | 42.5 |
| 12334 | 0.79 | 4.71 | | 1.15 | 7.4 | 32 | 145 | 0 | | 92 |
| 88937 | 0.42 | 2.45 | | 0.03 | 3.02 | 35 | 138 | 0 | | 44 |

3. **Research Study**

In this empirical study, we have collected data from four sections (01, 02, 04, and DL) of one course (IT 341) offered in Spring 2017 semester at GMU. Our data set includes various factors for each student, such as total time spent in hours on the week of the day, total logins, number of times and time spent on course content, homework assignments, lab sessions and skills assessment. The data set contains 59 factors and some of them are shown in Table 1.

We collected navigation behavioral data for each student until the midterm exam. The data from each student can be drilled down further in terms of number of times the student accessed the course content and how much time the student spent each day of the week as shown in Figures 1(a) and 1(b).

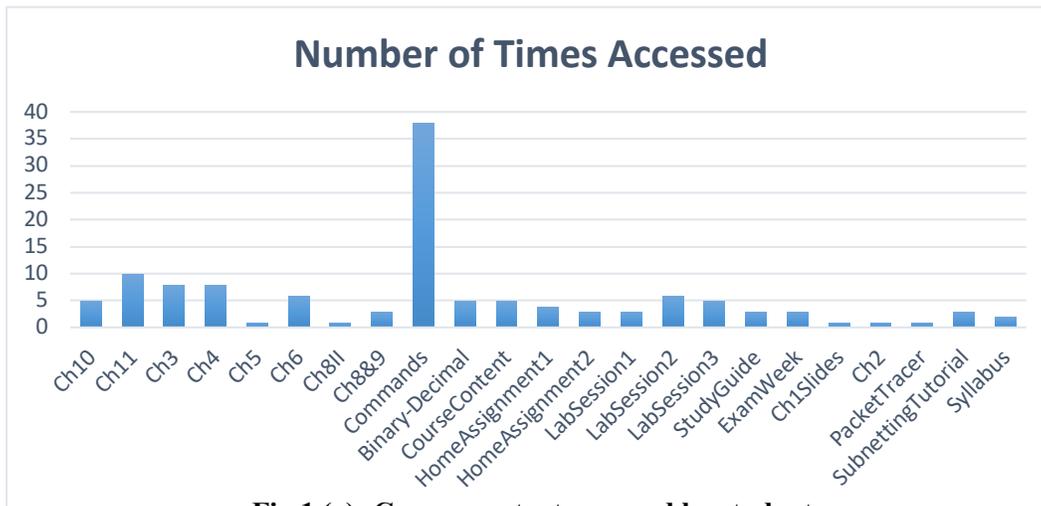

**Fig 1 (a): Course content accessed by student**

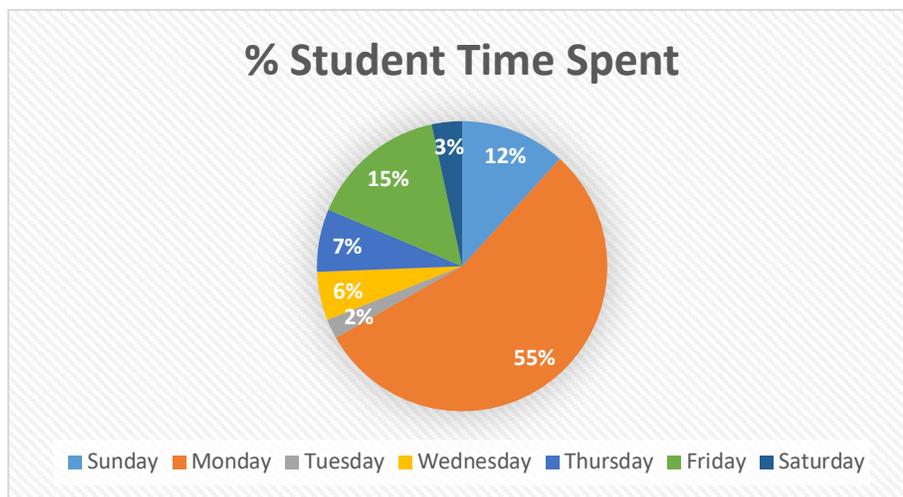

**Fig 1 (b): Percentage of the time spent by student**

The histogram of scores below show the distribution of scores for students on the scale of hundred. The histogram doesn't represent a normal distribution curve and shows that the most of the students' scores lies between 50 and 80 points. Furthermore more number of students are observed between 80 and 100 scores than between 30 and 50 scores.

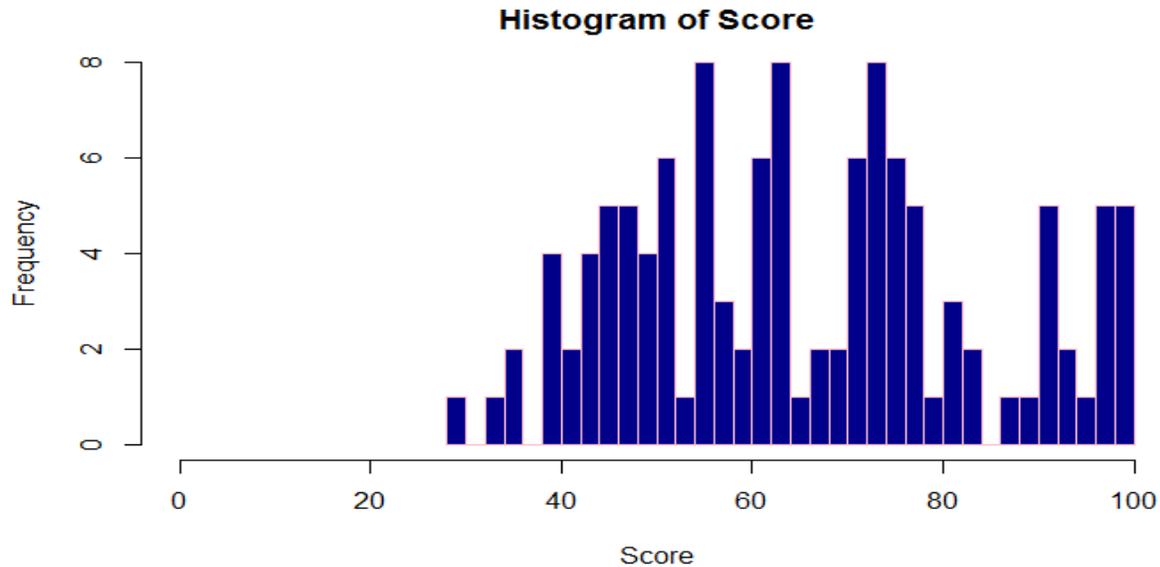

**Fig 2: Histogram Plot**

Data Preprocessing [11] is the first most important thing that we have to perform before building any model because it affects prediction performance and certain machine-learning algorithm that requires clean data. Since our dataset also contains some missing values for the students, we cannot simply assign missing values or take average to fill out the gaps. Therefore, we have removed such students to ensure clean and complete data for our modelling process. In exploratory data analysis, we observed that the time spent for accessing the course content is insignificant so we have only used number of times that the student accessed the course content. We have also observed that time spent on the day of the week varies across many students so it can be regarded as a good predictor to determine whether the student is disciplined, consistent and striving to perform well in the course.

Data Encoding [12] is an important aspect of predictive modelling and it has substantial influence on the models' performance. Hence, to model our data, we have encoded students' scores as either '0' or '1'. '0' represents the grade C/D/F and '1' represents the grade A/B. Now we have segmented the numerical scores in two different segments and we have built a classification model to predict the students' performance.

We have developed a model employing Random Forest [10] to predict students' performance. The main objective was to identify the key features that can predict students' performance with high accuracy.

4.  **Analysis and Results**

A. **Correlation Analysis**

The term correlation [13] weighs the relationship between the two variables. It is measured on the scale of '-1' to '+1'. In statistics, correlation coefficient measures the linear dependence between two variables. Correlation coefficient is calculated by dividing the covariance of the variables divided by the product of standard deviations of variables. If the correlation coefficient is equal to '0', it means that variables are not linearly related and there is no relationship between them. The magnitude of the correlation coefficient determines the strength of the relationship between the variables. If the correlation coefficient is less than zero, it means the variables have an inverse relationship, and if the correlation coefficient is greater than zero, it means the variables have a direct relationship.

We have plotted a correlation plot to investigate the relationship between the day of the week and students' scores. The day of the week represents the time spent in hours until the midterm, and score represents the midterm grade. The diagonal blue squares represent the correlation between identical pairs and it will always be '1' as shown in Figure 3. The blue color represents direct relationship, whereas red color represents inverse relationship. The correlation plot shows that 'Saturday' has the strongest direct relationship followed by 'Friday' and Monday'. We can also observe that Score has a negative relationship with Tuesday and positive relationship with other remaining days.

In other words, we can assume that if the student spends more time on the LMS on a Saturday, the student has a better chance of receiving a good score on the midterm exam. Other days represent a positive relationship with the score except Sunday and Tuesday. It means that if the student spends more time on these days then the student is less likely to score good on the exam.

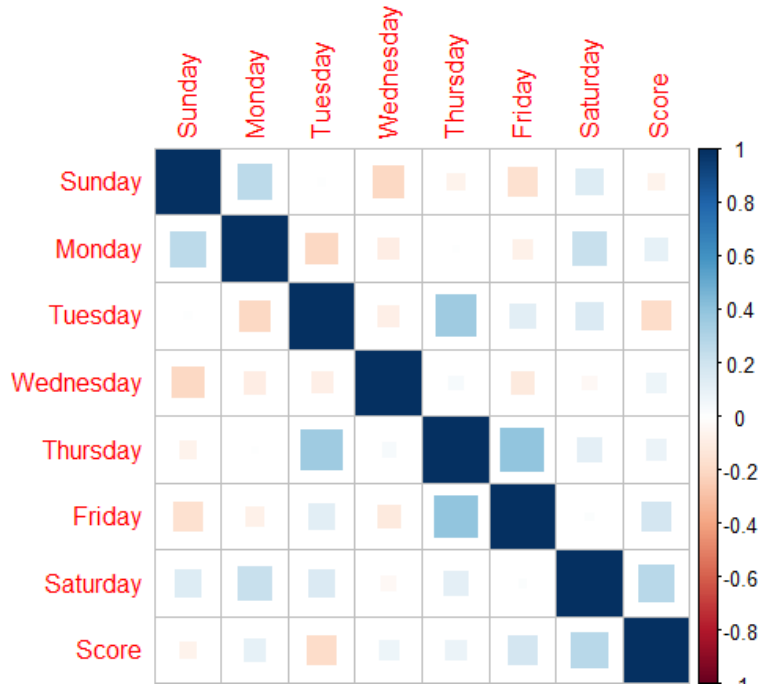

**Fig 3: Correlation plot for day of the week**

Furthermore, we need to develop a predictive model to study the interaction effect among the days of the week to predict the students' performance.

After observing the relationship of day of the week with the score, we investigated how course content and score are related. We have developed a correlation plot as shown in Figure 4. Here the course content such as 'Ch10Slides' mean the number of times the students accessed Chapter 10 Slides until the midterm. We can clearly identify that 'Ch4Slides' has the strongest relationship with Score followed by 'Ch6Slides', 'Ch5Slides' and 'Ch3Slides'. The main reason behind this relationship is that we designed the midterm exam based mostly on the concepts from these slides. Nonetheless, it requires deeper analysis to unravel more insights through the data.

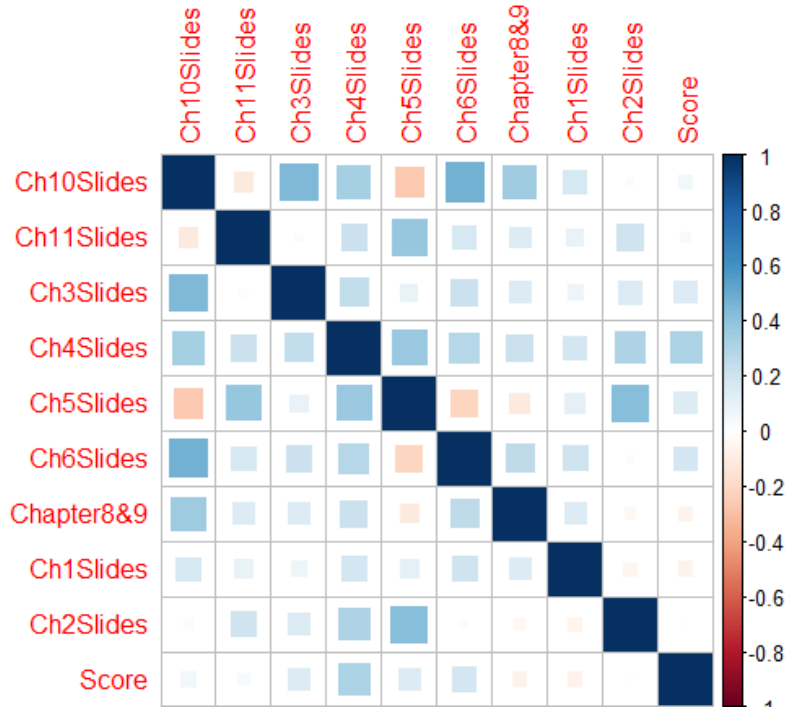

Fig 4: Correlation plot for Course Content

B. Predictive Model and Results

We are using Random Forest [10] as a machine-learning algorithm to identify the significant factors to predict the students' performance. Random Forest is one of the best choices because it decorrelates trees and avoids overfitting. In Random Forest, we build a number of decision trees on bootstrapped training samples. However, when building these trees, a random sample of subset of features is selected at each split in a tree.

For this model, we have used all the features present in the dataset to identify the significant features, which can help us to predict the students' performance. We have built a classification model of 500 decision trees with 11 variables selected at each split of the tree as shown in Figure 5 (a). The Random Forest results provide an accuracy of 77.78% for our model or the out-of-bag estimate [10] error rate of 22.22%.

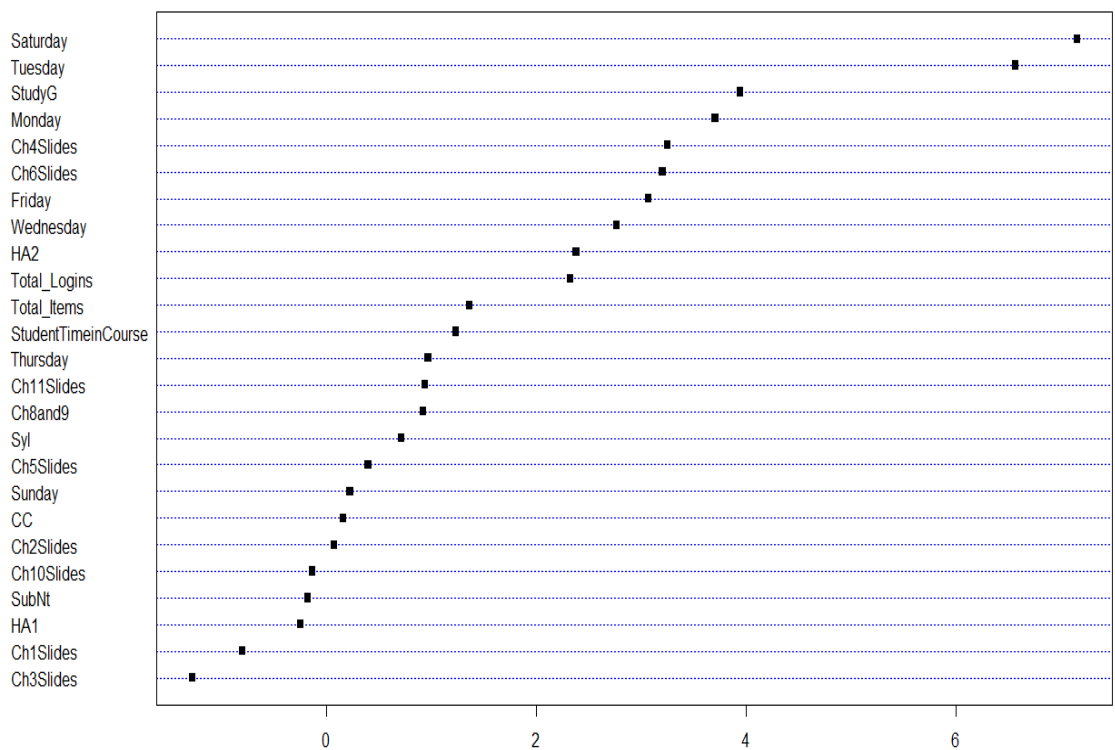

**Fig 5 (a): Output of complete model**

The graph shown below shows the order of the significant predictors. The most important predictor in our feature space is 'Saturday' followed by 'Tuesday', 'Study Guide', 'Monday', , Ch4Slides', 'Ch6Slides', 'Friday',etc.

**Fig 5 (b): Variable Importance Plot**

The variable importance plot shows the order of significant predictors in decreasing order. The variables in this plot are critical in grasping the performance of the student. For instance, study time spent on a Saturday is considered as the most important predictor, and if we exclude this data point from our model, it can significantly decrease the accuracy of our model. One possible explanation can be the home assignments or lab assignments that has to be submitted on a

Saturday. Students who submit their homework and lab assignments on time or students who study regularly every week tend to have a better chance of doing well in the midterm exam. Although Saturday is the best predictor, we still need to take into account factors such as other days of the week, course content and total logins of students in predicting their performance because these features have a synergy affect. The confusion matrix shown in Figure 5(a) represents '0' as C/D/F and '1' as A/B. which clearly shows that we are identifying the students who are going to earn C/D/F grade with an accuracy of 95.6% or classification error of 4.4%. This resonates with our target to identify students who are not going to perform well.

## 5. Conclusion and Future Works

We have accurately built a good classification model that identifies students who are not going to perform well on the midterm. Using this model, we can reach out to those students and provide guidance before the exam. Furthermore, it also helps the teachers to better manage their course content and put more emphasis to those lectures and supplementary materials that are significant to the midterm exam. Nonetheless, our data is still small and contains more '0' than '1' which means the classes are not evenly balanced We need to collect far more data to balance out the '0' and '1' in our data set and to scale our study. Furthermore, we can experiment by building two separate models to study the effect of the day of the week and the course content, and compare the accuracy of our models. This study can further be applied to other courses in order to help students perform better and manage course content effectively.

**Mr. Salman Yousaf,** *George Mason University*

Salman Yousaf is a graduate student in the field of Data Analytics Engineering. His research interests include learning analytics, big data, distributed machine learning and distance education for instructor and student learning. Salman earned his bachelor's degree at National University of Sciences and Technology and worked in the telecom industry in multiple roles such as Radio Optimization Engineer, Cell Planning Engineer and Technical Performance Analyst.

**Dr. Pouyan Ahmadi,** *George Mason University*

Pouyan Ahmadi is an Assistant Professor in the Department of Information Sciences and Technology. His research interests include cooperative communications and networking, cross-layer design of wireless networks, relay deployment and selection in wireless networks. In 2013, Dr. Ahmadi received the Best Graduate Student Paper Award at Wireless Telecommunications Symposium (WTS). He also won the Presidential Scholarship Award in 2010 at GMU. Dr. Ahmadi earned his Ph.D. in Electrical and Computer Engineering at George Mason University, M.S. in Architecture of Computer Systems at Iran University of Science and Technology, and B.S. in Computer Engineering at Azad University.

**Dr. Khondkar Islam,** *George Mason University*

Khondkar Islam is an Associate Professor and Associate Chair for Undergraduate Studies in the Department of Information Sciences and Technology. Dr. Islam is the Coordinator for the Networking Concentration. He is also the founder and director of the Virtual Academy (VirtAc) at George Mason University (GMU). His research interests include distributed and peer-to-peer systems, overlay and wireless networks, network security, and distance education for instructor training and student learning. Dr. Islam earned his Ph.D. and B.S. at George Mason University, and M.S. at American University.